\documentclass[twocolumn,reprint,aps,prb,showkeys,groupaddress,superscriptaddress,sectionbib,square,longbibliography]{revtex4-1}
\setcitestyle{numbers,square}
\usepackage{epsfig}
\usepackage{color}
\usepackage{times}
\usepackage{amsmath}
\usepackage{amssymb}
\usepackage{indentfirst}
\usepackage{graphicx}
\usepackage{subcaption}
\usepackage{bm}
\usepackage{wrapfig}

\newcommand{\fg}[1]{\mbox{\pmb{$#1$}}}
\newcommand{\bey}{\begin{eqnarray}}
\newcommand{\eey}{\end{eqnarray}}
\newcommand{\vep}{\varepsilon}
\newcommand{\fvep}{\fg \varepsilon}
\newcommand{\fsg}{\fg \sigma}
\newcommand{\sg}{\sigma}
\newcommand{\bec}{\begin{center}}
\newcommand{\eec}{\end{center}}

\newcommand{\ssst}{\scriptscriptstyle}

\begin{document}
\title{Deformation, lattice instability, and metallization during solid-solid structural transformations under general applied stress tensor:  example of Si I $\rightarrow$ Si II  }

\author{Nikolai~A. Zarkevich}\email{zarkev@ameslab.gov}
\affiliation{Ames Laboratory, U.S. Department of Energy, Iowa State University, Ames, Iowa 50011-3020, USA}

\author{Hao Chen}\email{haochen@iastate.edu}
\affiliation{Department of Aerospace Engineering, Iowa State University, Ames, Iowa 50011, USA}

\author{Valery I. Levitas}\email{vlevitas@iastate.edu}
\affiliation{Ames Laboratory, U.S. Department of Energy, Iowa State University, Ames, Iowa 50011-3020, USA}
\affiliation{Department of Aerospace Engineering, Iowa State University, Ames, Iowa 50011, USA}
\affiliation{Department of Mechanical Engineering, Iowa State University, Ames, Iowa 50011, USA}
\affiliation{Department of Materials Science \& Engineering, Iowa State University, Ames, Iowa 50011, USA}

\author{Duane~D. Johnson}\email{ddj@iastate.edu, ddj@ameslab.gov}
\affiliation{Ames Laboratory, U.S. Department of Energy, Iowa State University, Ames, Iowa 50011-3020, USA}
\affiliation{Department of Materials Science \& Engineering, Iowa State University, Ames, Iowa 50011, USA}

\date{\today}

\begin{abstract}
Density functional theory (DFT) was employed to study the stress-strain behavior,  elastic instabilities,
and metallization during a solid-solid phase transformation (PT) between semiconducting Si I (cubic A4) and metallic Si II (tetragonal A5 structure) when subjected to a {\it general stress tensor}.
With normal stresses ($\sg_1$, $\sg_2$, $\sg_3$) acting along  $\left<110 \right>$,  $\left<1\bar{1}0 \right>$, and $\left<001\right>$, respectively, dictating the simulation cell,
we determine combinations of 6 independent stresses that drive a lattice instability for the Si I$\rightarrow$Si II PT, and a semiconductor-metal electronic transition.
Metallization precedes the structural PT, hence, a stressed Si I can be a metal.
Surprisingly, a stress-free Si II is metastable in DFT.
Notably, the PT for hydrostatic pressures is at 75.81 GPa,  while under uniaxial stress it is 11.03 GPa (or 3.68 GPa mean pressure).
Our key result: The Si I $\rightarrow$ Si II PT is described by a critical value of the modified transformation work, as found with a phase-field method,
and the PT criterion  has only two parameters for a general applied stress.
More generally, our findings are crucial for revealing novel (and more economic) material synthesis routes for new or known high-pressure phases under controlled and predictable non-hydrostatic loading and plastic deformation.
\end{abstract}

\maketitle

\section{\label{INTRODUCTION}INTRODUCTION}

Phase transformations (PTs) in solids are traditionally characterized by temperature-pressure ($T$-$P$) phase diagrams at thermodynamic equilibrium \cite{Tonkov-Ponyatovsky-2004}, whereas general non-hydrostatic  stresses offer novel synthetic routes for new or known high-pressure phases. Here, we augment standard $T$-$P$ diagrams of structural (and electronic) instabilities, providing guidance for creating more accessible processing routes of such phases under controlled and predictable non-hydrostatic deformation at significantly lower mean pressures.
Indeed, observed PTs occur under significant deviation from equilibrium \cite{PRB91p174104y2015,Schindler-Vohra-1995,Solozhenko-1995,Ji-etal-12}, and most first-order PT exhibit a hysteresis. For carbon, the equilibrium pressure for graphite-diamond PT at room temperature is 2.45 GPa; however, due to hysteresis, the PT is observed to start at 70 GPa \cite{Schindler-Vohra-1995}.
The high-pressure superhard cubic boron nitride (BN) is stable at ambient conditions \cite{Solozhenko-1995},
however, disordered hexagonal BN does not transform even at 52.8 GPa \cite{Ji-etal-12}.
The actual PT pressure deviates from that of equilibrium due to an enthalpy barrier.
When thermal fluctuations can be neglected, e.g., at low temperature and short times,
the PT criterion is related to disappearance of the enthalpy barrier, i.e., to the lattice instability.
Hence, lattice instability conditions are intensively studied under hydrostatic, uniaxial, and multi\-axial loadings \cite{04,Milstein-PRL-95,Wang-PRL-93,Grimvall-etal-2012,Pokluda-etal-2015,Levitasetal-Instab-17,Levitasetal-PRL-17}.
While phase equilibrium  under stress tensor can be derived within continuum thermodynamic treatment for elastic \cite{Grinfeld-1991} and elastoplastic \cite{Levitas-IJSS-1998} materials, lattice instabilities require a separate consideration.

{\par} In experiments, there is a significant reduction in the PT pressure due to the  deviatoric (non-hydrostatic) stresses and plastic strains \cite{levitas-prb-04,Blank-Estrin-2014,Ji-etal-12,Edalati-Horita-16,Levitas-JPCM-18}.
For example, a large plastic shear reduces the PT from highly disordered to superhard wurtzitic BN from 52.8 to 6.7 GPa  \cite{Ji-etal-12} -- an order of magnitude!
This phenomenon is extremely important from fundamental and applied points of view, as it may reduce the PT pressure to a practical level for high-hydrostatic-pressure phases that exhibit unique properties.

{\par}The suggested physical mechanism responsible for this reduction is related to dislocation pileups associated with a plastic strain \cite{levitas-prb-04}.
As stresses at the tip of a pileup are proportional to the number of dislocations in a pileup (typically 10 to 100), local stresses exceed the lattice instability limit and cause nucleation of a high-pressure phase even at relatively small external pressure.
This was rationalized based on an analytical model \cite{levitas-prb-04} and using a phase field approach \cite{Levitas-Mahdi-Nanoscale-14,Javanbakht-Levitas-PRB-16}.
However, the phase field inputs for the forward (direct) and reverse PT instability criteria for an ideal crystal under {\it general stress tensor} was assumed hypothetically, as such criteria are not known for any material.
In addition, for many materials there is a significant difference between calculated instability pressure (e.g., 64-80 GPa for Si I$\rightarrow$Si II PT \cite{yip-prb-hydrostatic,Gaal-Nagy-etal-2001}) and experimentally determined PT pressure (e.g., 9-12 GPa for the same PT \cite{Domnichetal-04}). This reduction was qualitatively explained by presence of the local stress concentrators around defects (dislocations, grain boundaries,  etc.) and the effect of the non-hydrostatic stresses. Quantitative solution of this problem requires knowledge of the lattice instability conditions for a given stress tensor.

Notably, due to the technological importance of Si and its PTs, a huge literature exists.
Relevant are the PTs in Si I under hydrostatic and two-parametric nonhydrostatic loadings,
studied with DFT  \cite{yip-prb-hydrostatic,Gaal-Nagy-etal-2001},
and the lattice instability under  two-parametric  nonhydrostatic loadings (unrelated to a PT) \cite{Umeno-Cerny-PRB-08,Cernyetal-JPCM-13,Telyatniketal-16,Pokluda-etal-2015}.
Importantly, PT in Si under plastic deformations is utilized in the ductile regime for machining \cite{Patten-04}.

{\par} Thus, we perform a DFT-based study of the deformation process under applied {\it general stress}.
We determine the lattice instabilities, responsible for the cubic-to-tetragonal Si~I$\rightarrow$Si~II PT, and metallization.
While finding the instability criteria under all six stress components seems daunting,  due to the large number of combinations,
an unexpected guidance came from the
crystal lattice instability criterion formulated within the phase-field method   \cite{Levitas-IJP-13,Levitasetal-Instab-17,Levitasetal-PRL-17,Levitas-IJP-18}.
The key result is that
Si I$\rightarrow$Si II PT can be  described by the  critical value of the modified transformation work.
With normal stress $\sg_3$ in  $\left<001\right>$ direction and $\sg_1$ and $\sg_2$ acting along  $\left<110 \right>$ and  $\left<1\bar{1}0 \right>$, respectively,
the PT criterion is linear in normal stresses, depends on $\sg_1+\sg_2$, is independent of $\sg_1-\sg_2$ and shear stress $\tau_{21}$, acting alone or with one more shear stress,
and depends on all shear stresses via theoretically predicted  geometric nonlinearity with zero linear term.
 The PT criterion has only two material parameters for a general applied stress, which can be determined by two DFT simulations under different normal stresses.

\section{\label{METHOD}SIMULATION METHODS}

We used DFT as implemented in VASP \cite{VASP1,VASP2,VASP3} with the projector augmented waves (PAW) basis \cite{PAW,PAW2} and PBE exchange-correlation functional \cite{PBE}.
The PAW-PBE pseudo-potential of Si had 4 valence electrons ($s^2 p^2$) and 1.9 {\AA} cutoff radius.
The plane-wave energy cutoff (ENCUT) was 306.7~eV, while an augmentation charge (ENAUG) was 322.1~eV.
We used a Davidson block iteration scheme (IALGO=38) for the electronic energy minimization.
Electronic structure was calculated with a fixed number of bands (NBANDS=16) in a tetragonal 4-atom unit cell (a supercell of a 2-atom primitive cell).
Brillouin zone integrations were done in $k$-space (LREAL=FALSE), using a $\Gamma$-centered Monkhorst-Pack mesh \cite{MonkhorstPack1976} containing 55 to 110 $k$-points per {\AA$^{-1}$} (fewer during atomic relaxation, more for the final energy calculation).
Accelerated convergence of the self-consistent charge calculations was achieved using a modified Broyden's method \cite{PRB38p12807y1988}.

{\par} Atomic relaxation in a fixed unit cell (ISIF=2) was performed using the conjugate gradient algorithm (IBRION=2), allowing symmetry breaking (ISYM=0).
The transformation path was confirmed by the nudged elastic band (NEB) calculations, performed using the C2NEB code \cite{C2NEB}.
We used DFT forces in \emph{ab initio} molecular dynamics (MD) to verify stability of the relaxed atomic structures.
Si atoms were assumed to have mass POMASS=28.085 atomic mass units (amu).
The time step for the atomic motion was set to POTIM=0.5 fs.
Additionally,  our classical MD simulations used a Tersoff  potential (TP), as described in \cite{tersoff}.

\section{\label{LATTICE} ENERGY LANDSCAPE} 

\begin{figure}[t]
    \includegraphics[width=0.4\textwidth]{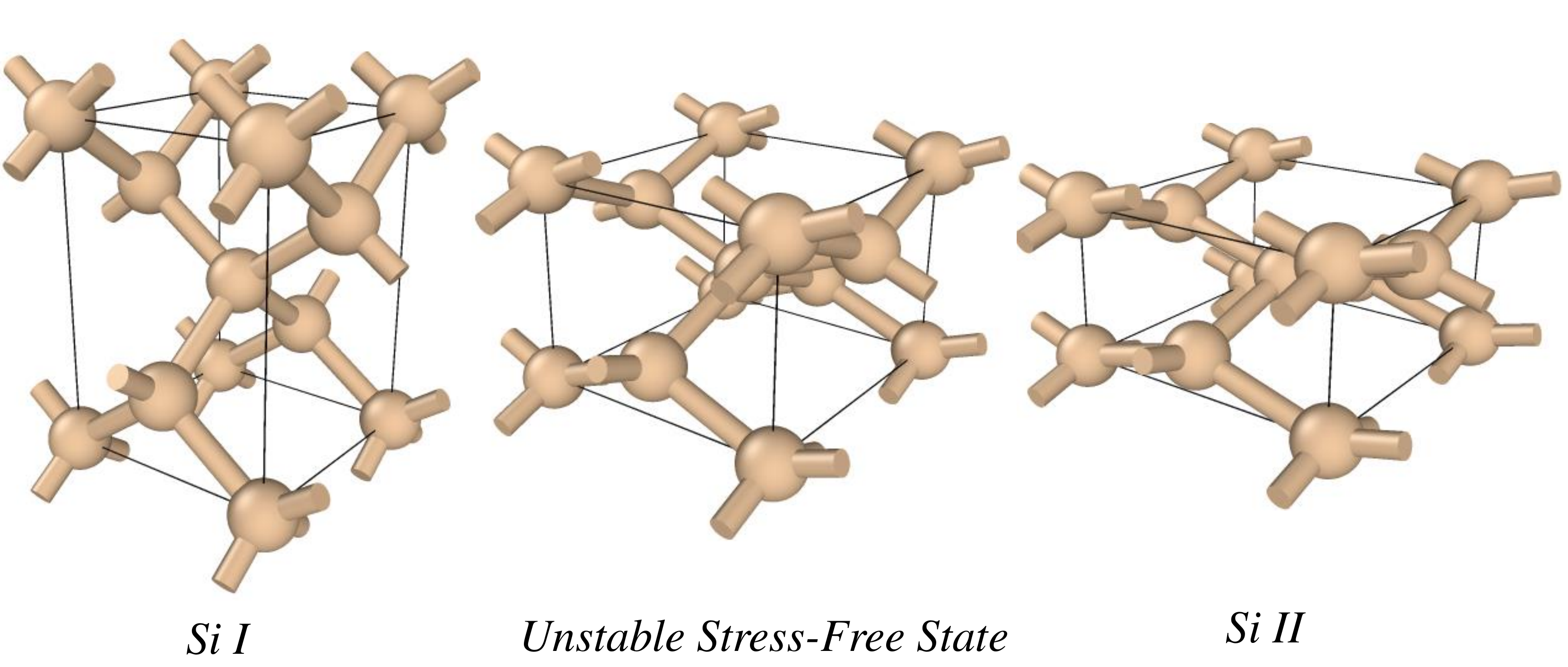}
    \caption{Si  atoms and nearest-neighbor bonds in tetragonal ($a=b$, $c$) non-primitive unit cells are given for stress-free Si I (left), unstable SP (middle),
    and Si II (right).
      }
    \label{Fig1}
\end{figure}

The calculated potential energy (versus the tetragonal lattice constants $c$ and $a\!=\!b$) is shown in Fig.~\ref{Fig2}.
All the primary data of our simulations and data for each figure are placed in Supplemental Material \cite{suppl}.
Using DFT, we found two local energy minima, corresponding to the fully relaxed stress-free Si I and Si II,
and a saddle point (SP) -- an intermediate unstable state corresponding to the enthalpy barrier, see Fig.~\ref{Fig1}.
The tetragonal cell of Si I is bounded by  $(110)$,  $(1\bar{1}0)$, and  $(001)$ planes.
The calculated
energies (and tetragonal lattice constants $a_i=b_i$ and  $c_i $) relative to the stress-free Si I ($a_1 =3.8653\,${\AA}, $c_1=\sqrt{2}a_1 =5.4665 \,${\AA})
are $0.2949~$eV/atom for Si II  ($a_2 \! = \!4.8030 \,${\AA}, $c_2 \! = \! 2.6592 \,${\AA}), and
 $0.4192~$eV/atom for the SP state  ($a \! = \! 4.4847 \,${\AA}, $c \! = \! 3.4763 \,${\AA}).
The calculated $c_1$ 
 is within 0.7\%  of experiment ($5.43\,${\AA}) \cite{Yim-Paff-74}.

\begin{figure}[b]
    \includegraphics[width=0.4\textwidth]{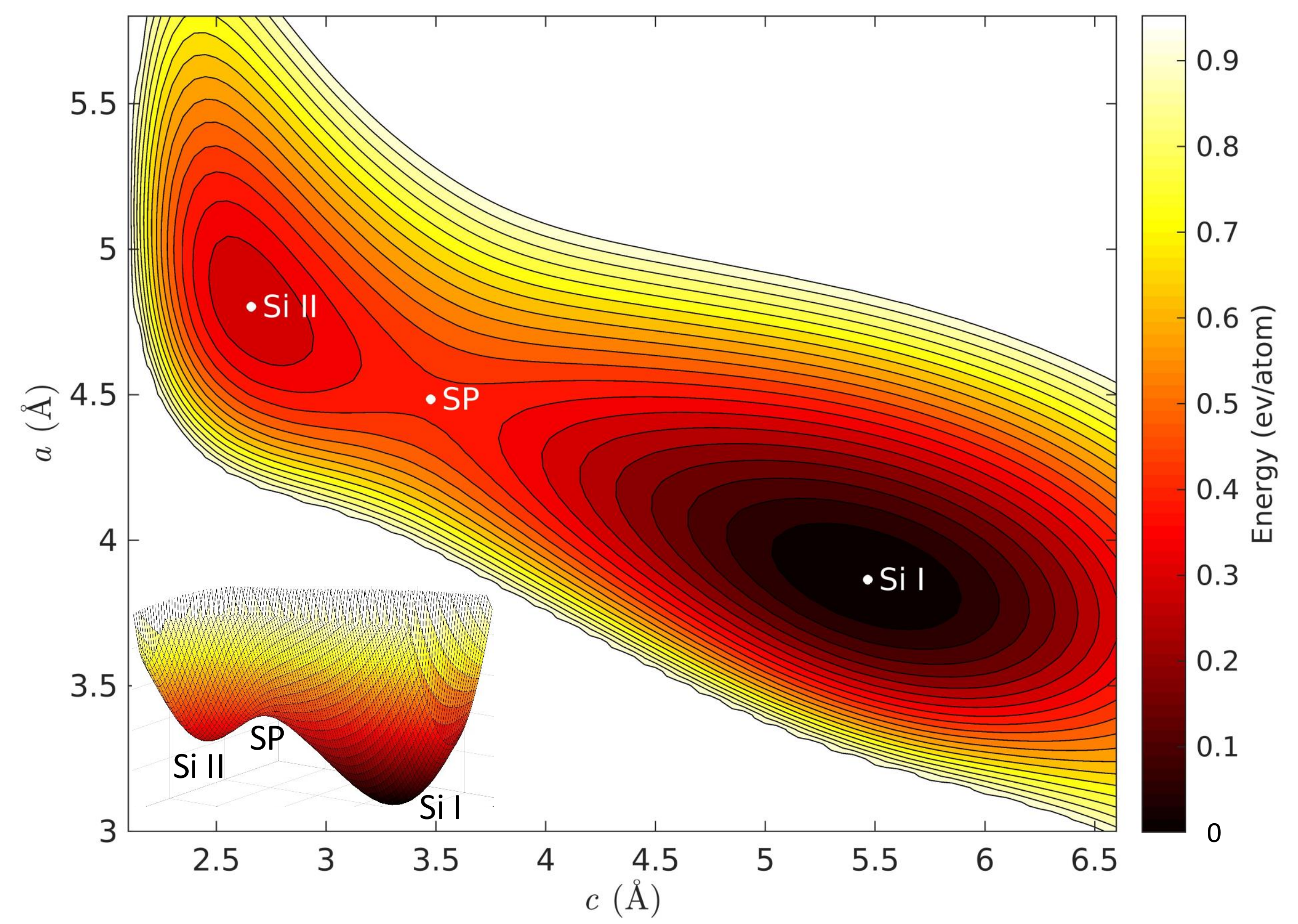}
    \caption{ DFT energy of Si versus lattice parameters $c$ and $a$.
      }
    \label{Fig2}
  \end{figure}

\section{\label{SSCURVES}STRESS-STRAIN CURVES} 

\begin{figure*}
    \includegraphics[width=0.40\textwidth]{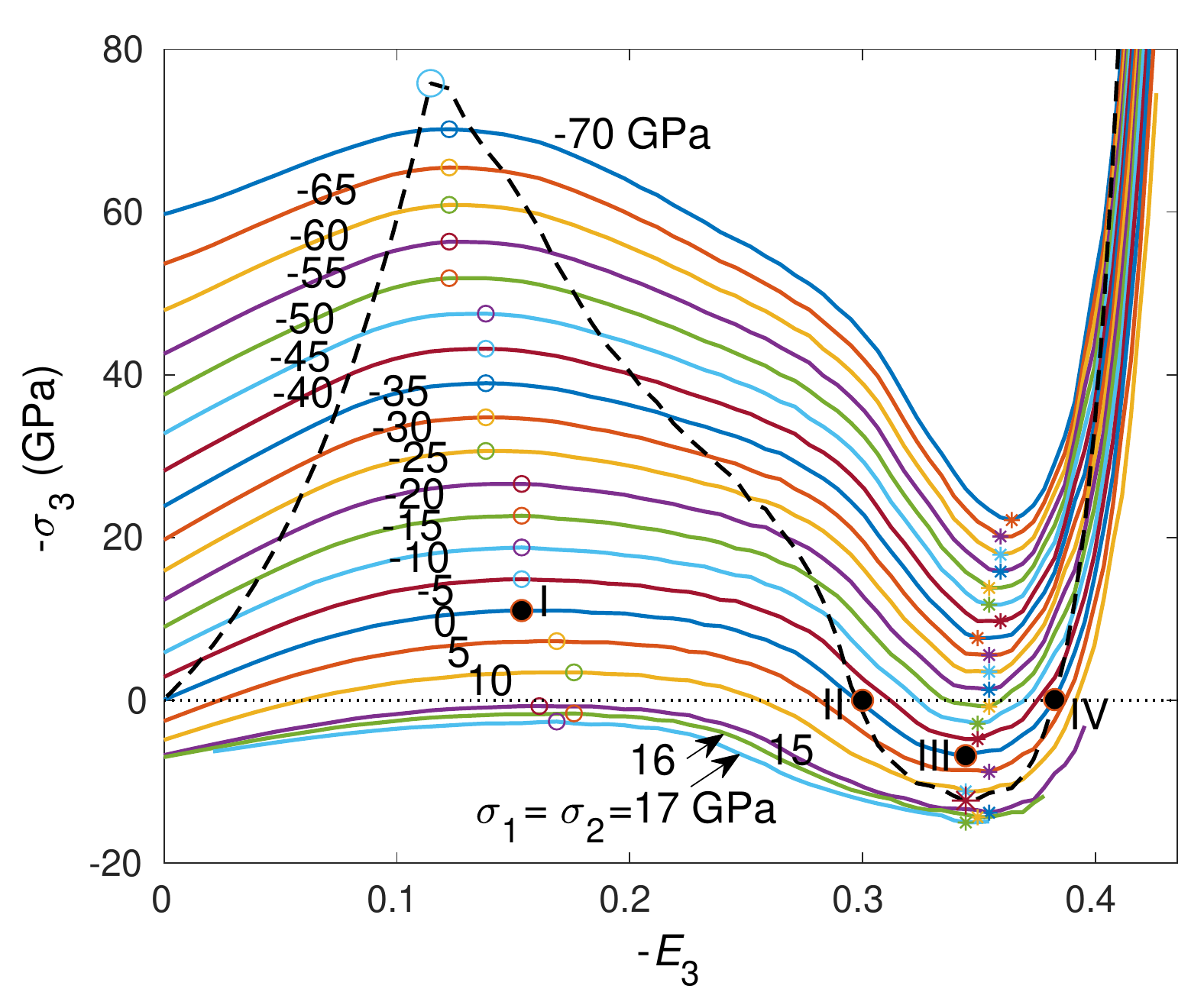}  
    \includegraphics[width=0.40\textwidth]{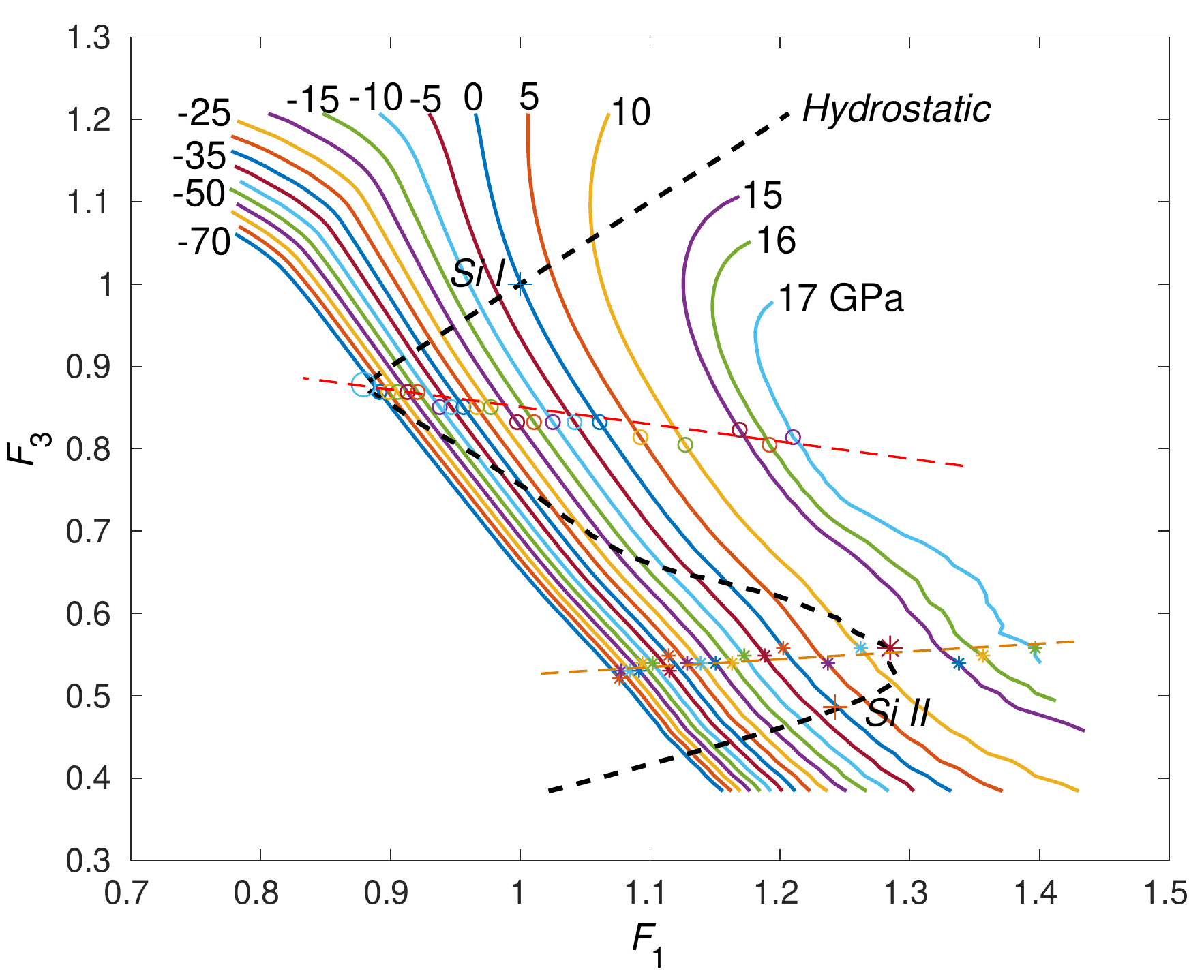}  
    \caption{\label{Fig3} (a) True (Cauchy) stress versus Lagrangian strain, i.e., $\sg_3$--$E_3$  curves, for compression/tension
    along $c$ for various lateral $\sg_1=\sg_2$ stresses for Si I$\leftrightarrow$Si II PTs and (b) corresponding transformation paths in $(F_1 \! = \! F_2, F_3)$ plane.  Hollow (solid) symbols mark instability points for forward (reverse) PT. Stress-strain curve (dashed line) for hydrostatic loading is included.
      }
\end{figure*}

{\par} Tensors are designated with boldface. 
${\fg I}$ is the unit tensor.
Contractions of tensors ${\fg A}=\{A_{ij}\}$ and ${\fg B}=\{B_{jk}\}$ over one and two indices  in Einstein notations are
 ${\fg A}{\fg \cdot}{\fg B}=\{A_{ij}\,B_{jk}\}$ and  ${\fg A}{\fg :}{\fg B}=A_{ij}\,B_{ji}$, respectively.
The inverse and transpose of ${\fg A}$ are ${\fg A}^{-1}$ and ${\fg A}^{\ssst T}$, respectively.

{\par}  The deformation gradient $\fg F=\fg F_e \cdot \fg U_t$, which maps initial undeformed state of a crystal into current deformed state, can be multiplicatively decomposed  into elastic $\fg F_e$ and transformational $\fg U_t$ parts.
Transformation deformation gradient $\fg U_t$ changes the Si I stress-free cell to the Si II stress-free cell;
its diagonal components are $U_{t1}=U_{t2}= a_2/a_1=1.243$ and $U_{t3}=c_2/c_1=0.486$.
For comparison,  the Tersoff potential in \cite{Levitasetal-Instab-17,Levitasetal-PRL-17} leads to $U_{t1}=U_{t2}= 1.175$ and $U_{t3}=0.553$.

{\par} We use true Cauchy stress $\fsg$ (force per unit deformed area) and Lagrangian strain  ${\fg E} =\frac{1}{2}({\fg F}^{\ssst T} \cdot {\fg F}- \fg I)$.
Stress-strain curves $\sg_3$--$E_3$ in $c$ direction for fixed lateral stresses $\sg_1=\sg_2$ in $a$ and $b$ directions
are presented in Fig.~\ref{Fig3}, along with corresponding transformation paths in $(F_1 \! = \! F_2, F_3)$ plane.
The elastic instability occurs when determinant of the matrix of the elastic moduli, modified by some geometrically nonlinear terms, reduces to zero  \cite{04, Milstein-PRL-95, Wang-PRL-93, Grimvall-etal-2012, Pokluda-etal-2015}.
This results in a condition that some elastic moduli (or combination thereof) reduce to zero.
We will use an alternative condition, based on the following strict definition.
Elastic lattice instability at true stress $\fsg$ occurs at stresses above  (or below for the reverse PT) which the crystal cannot be at equilibrium.
The instability points correspond to the stress maximum for forward PT (and minimum for reverse PT) on the stress-strain curves, see Fig.~\ref{Fig3}.
Both conditions coincide in many cases, although they are not strictly equivalent.
The second condition is more general and universal, because it is applicable even to the cases with discontinuous or undefined derivatives
of stress with respect to strain. In those  cases the elastic moduli are not well-defined, and determinant of the matrix of the elastic moduli cannot be found.

{\par}  In Fig.~\ref{Fig3}, a tetragonal stressed lattice of Si I transforms into a tetragonal stressed lattice of Si II,
and the lattice instability does not change this tetragonal symmetry.
The slope of the stress-strain curve is continuous and is zero at instability points.
Under hydrostatic loading (dashed line in Fig.~\ref{Fig3}),  a cubic lattice looses its stability under tetragonal perturbations, i.e., there is a bifurcation from a primary isotropic  deformation to a secondary tetragonal deformation; hence, the derivative at the hydrostatic instability point is discontinuous.
Both under hydrostatic ($\sg_1=\sg_2=\sg_3$) and uniaxial ($\sg_1=\sg_2=0$) compression there are three stress-free states (Fig.~\ref{Fig1}):
 Si I, Si II (stable or metastable enthalpy minima) and an intermediate unstable state at the SP (enthalpy barrier).
Interestingly, a stress-free Si II is metastable, with stable phonons \cite{gaal2001ab,gaal2006phonons,phononSi18}.
Thus, one could search for a pressure-plastic shear path for arresting the metastable Si II, as suggested in  \cite{levitas-prb-04} for any metastable phase.
In theory, a stress-free cubic Si I is a deep global energy minimum,
while a stress-free tetragonal Si II is a shallow local energy minimum ($0.2949~$eV/atom above Si I and $0.1243~$eV/atom below SP),
which can be further destabilized by an internal stress that appear during forward martensitic PT.
In experiments, a stress-free Si II was not observed, while a
depressurized Si II does not reverse to Si I, but  transforms to Si XII and then to Si III under slow unloading, or to amorphous Si under fast unloading \cite{Domnichetal-04}.

\begin{figure}[b]
    \includegraphics[width=0.40\textwidth]{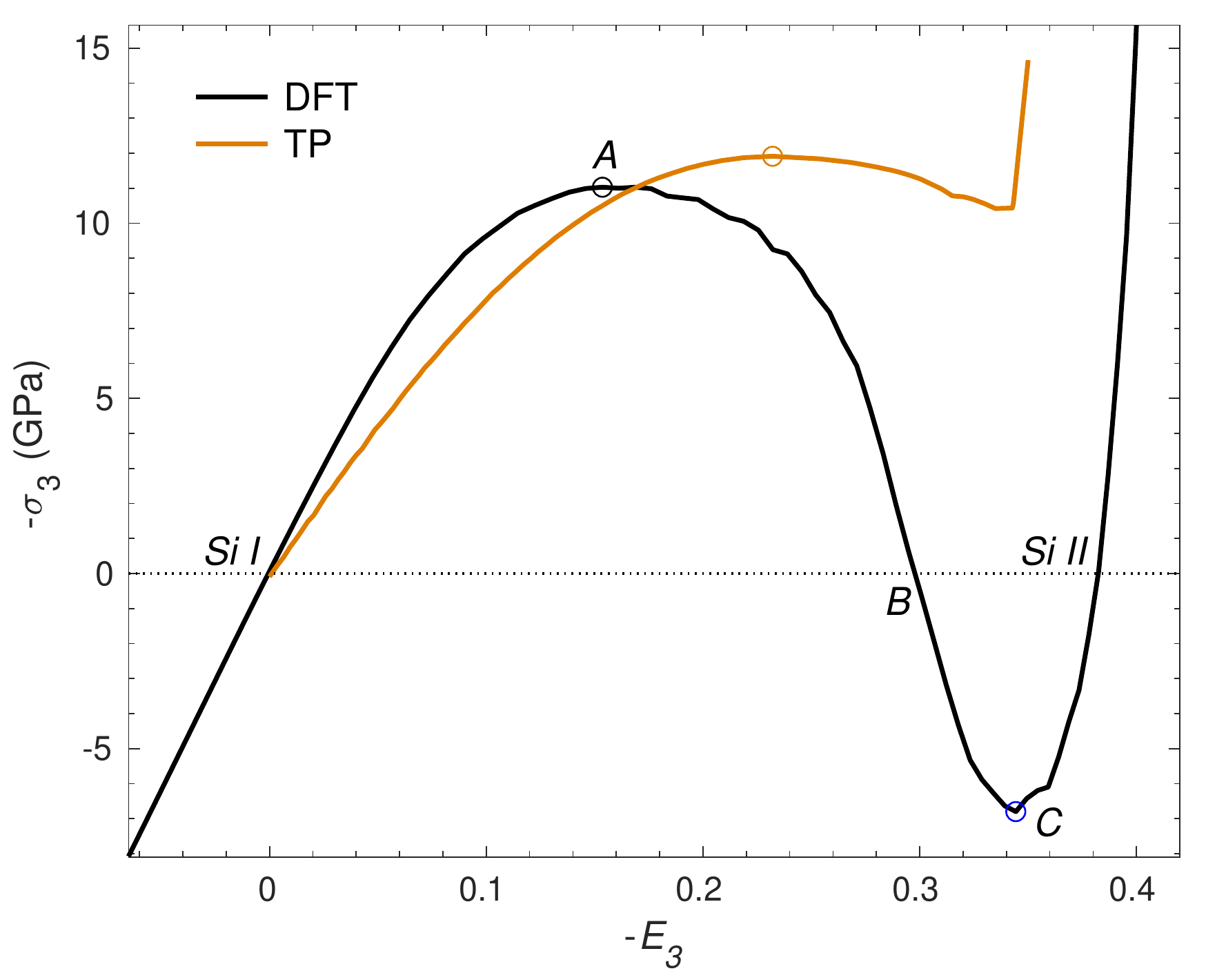}  
    \caption{ Comparison of  the Cauchy stress  vs. Lagrangian strain ($\sg_3$--$E_3$  curves) for a uniaxial compression
    along $c$ at fixed lateral stresses $\sg_1=\sg_2=0$ for Si I$\leftrightarrow$Si II PTs,  obtained from  DFT and Tersoff-based simulations.
      }
    \label{Fig4}
\end{figure}

{\par}Stress-strain $\sg_3$--$E_3$  curves for a uniaxial compression at $\sg_1=\sg_2=0$,  obtained with DFT and 
Tersoff potential, are compared in Fig.~\ref{Fig4}.
While the maximal stresses for Si I corresponding to the elastic lattice instability (see below) are comparable in both approaches,
other features differ significantly, including elastic rule for Si I, strain for the lattice instability of Si I, and the transformation strain.
The TP-based stress-strain curve does not intersect zero-stress axis, i.e., stress-free Si II is unstable in a classical force field.
The same is true for the Stillinger-Weber and modified-Tersoff  potentials \cite{Levitasetal-PRL-17}.

\section{\label{ELASTIC}ELASTIC LATTICE INSTABILITY UNDER TWO-PARAMETRIC LOADING}
\begin{figure}[b]
 \centering
 \includegraphics[width=0.4\textwidth]{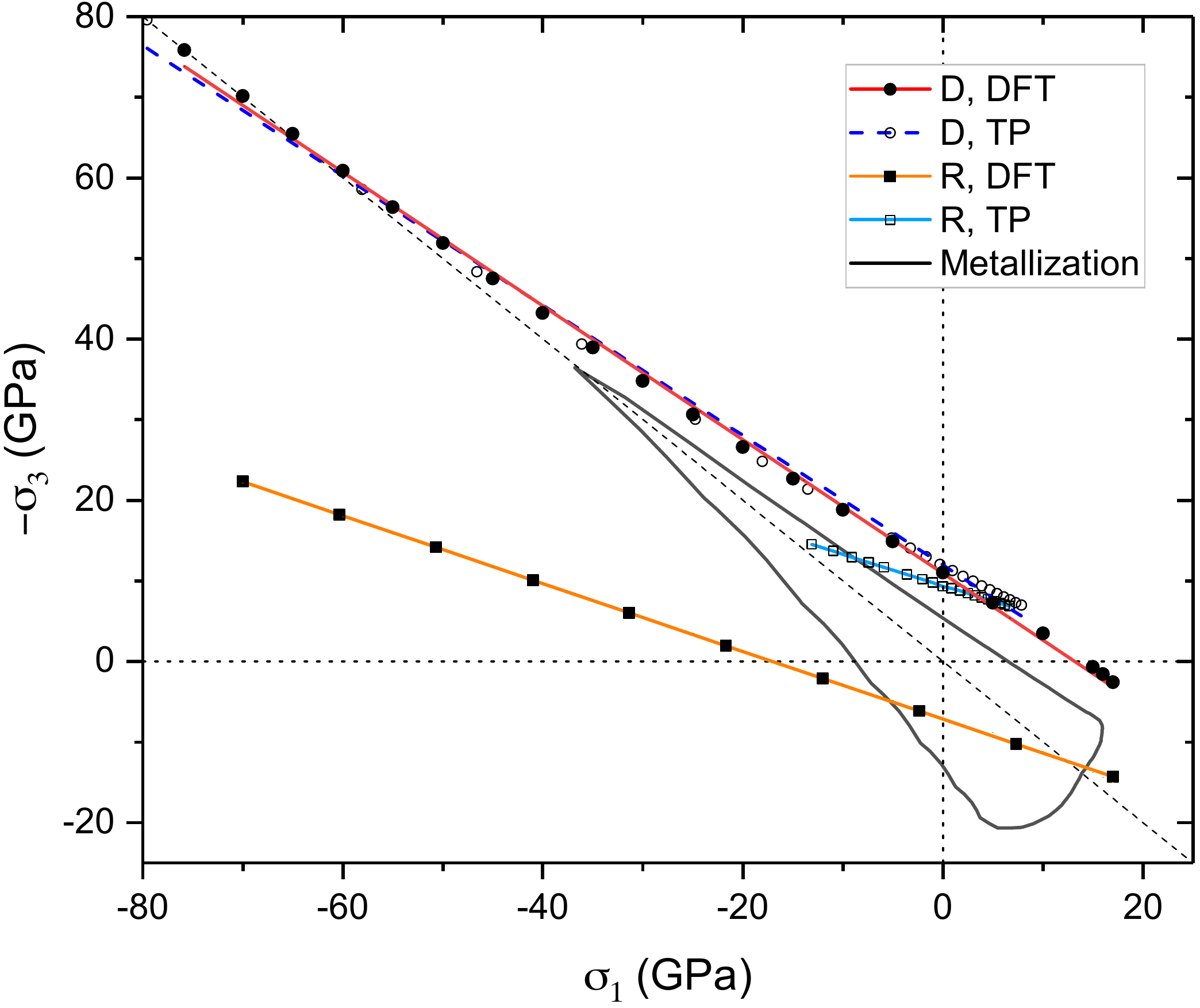}
    \caption{\label{Fig5}  Elastic instability vs. $\sg_3$ and $\sg_1=\sg_2$ for direct (D) Si I$ \rightarrow$Si II and reverse (R)  Si II$ \rightarrow$Si I PTs
    from DFT and TP-based results \cite{Levitasetal-Instab-17,Levitasetal-PRL-17},  and metallization curve from DFT.
The hydrostatic condition ($\sg_1 = \sg_2 = \sg_3$) is a diagonal (dashed black) line.
}
\end{figure}

Elastic lattice instabilities at $\sg_{1} \! = \! \sg_2$ for direct ($\sg_{3d}$) and reverse ($\sg_{3r}$) PTs are shown in Fig.~\ref{Fig5}.
Both instability conditions are approximated by linear relationships.
Tersoff-potential results from  \cite{Levitasetal-Instab-17,Levitasetal-PRL-17} for Si I$\rightarrow$Si II PT are generally in good agreement with the present DFT results,
however there is a difference for tensile and small compressive $\sg_1$, where TP results are slightly higher and nonlinear; at a tensile stress $\sg_1>8 $~GPa, they cross the instability line for the reverse PT.
Also, under hydrostatic loading, PT pressure from DFT and TP is 75.81~GPa and 79.58~GPa, respectively.

For uniaxial compression, the PT stress is $\sg_{3d}=- 11.03\,$GPa  at $E_3=-0.154 $.  In comparison, DFT simulations in \cite{Cernyetal-JPCM-13} give $\sg_{3d}=-10.6 \,$GPa and $E_3=-0.16$ (recalculated from engineering strain in \cite{Cernyetal-JPCM-13}); DFT results in  \cite{Telyatniketal-16} using two different methods suggest $\sg_{3d}=-11.9 \; (-12.7) $~GPa and $E_3=-0.14\; (-0.16)$ (recalculated from logarithmic strain in \cite{Telyatniketal-16});  and TP in \cite{Levitasetal-Instab-17,Levitasetal-PRL-17} gives  $\sg_{3d}=-12.03 \,$GPa  and $E_3=-0.232 $.
Note that the pressure for uniaxial loading is $-\sg_{3d}/3= 3.68 \,$GPa, which is $75.81/3.68=20.6$ times lower than under hydrostatic conditions. This characterizes very strong effect of non-hydrostatic stresses on PT pressure, which can partially explain significantly lower  experimental PT pressure than the instability pressure and scatter in experimental data under quasi-hydrostatic conditions.
The instability lines are described by
$\sg_{3d}= -10.9 + 1.20\sg_1$ for $\sg_1 \subset [-75.81; 17]$  and $\sg_{3r}= 7.175 + 0.4209 \sg_1$ for $\sg_1 \subset [-70; 17]$.
Theoretical strength in \cite{Cernyetal-JPCM-13} is approximated as $\sg_{3d}= -10.6 + 0.77 \sg_1$ for $\sg_1 \subset [-15; 12]$.
As it is close to our result, instability  in \cite{Cernyetal-JPCM-13} is related to Si I$ \rightarrow$Si II PT.

 While instability line for forward PT in \cite{Levitasetal-Instab-17,Levitasetal-PRL-17} with TP-MD is quite close to our DFT results, for reverse PT the TP-MD results are completely different from DFT.
 Consequently none of the classical potentials in \cite{Levitasetal-Instab-17,Levitasetal-PRL-17} (Tersoff, modified Tersoff, and Stillinger-Weber) are able to describe the reverse PT. This also means that phenomena related to coincidence of the forward and reverse PTs in some tensile lateral stress range predicted in \cite{Levitasetal-PRL-17} are not realistic for Si I. Still, they may be found in other materials.


\section{\label{METALLIZATION}METALLIZATION  UNDER BIAXIAL LOADING}

\begin{figure}[b]
 \centering
  \includegraphics[width=0.4\textwidth]{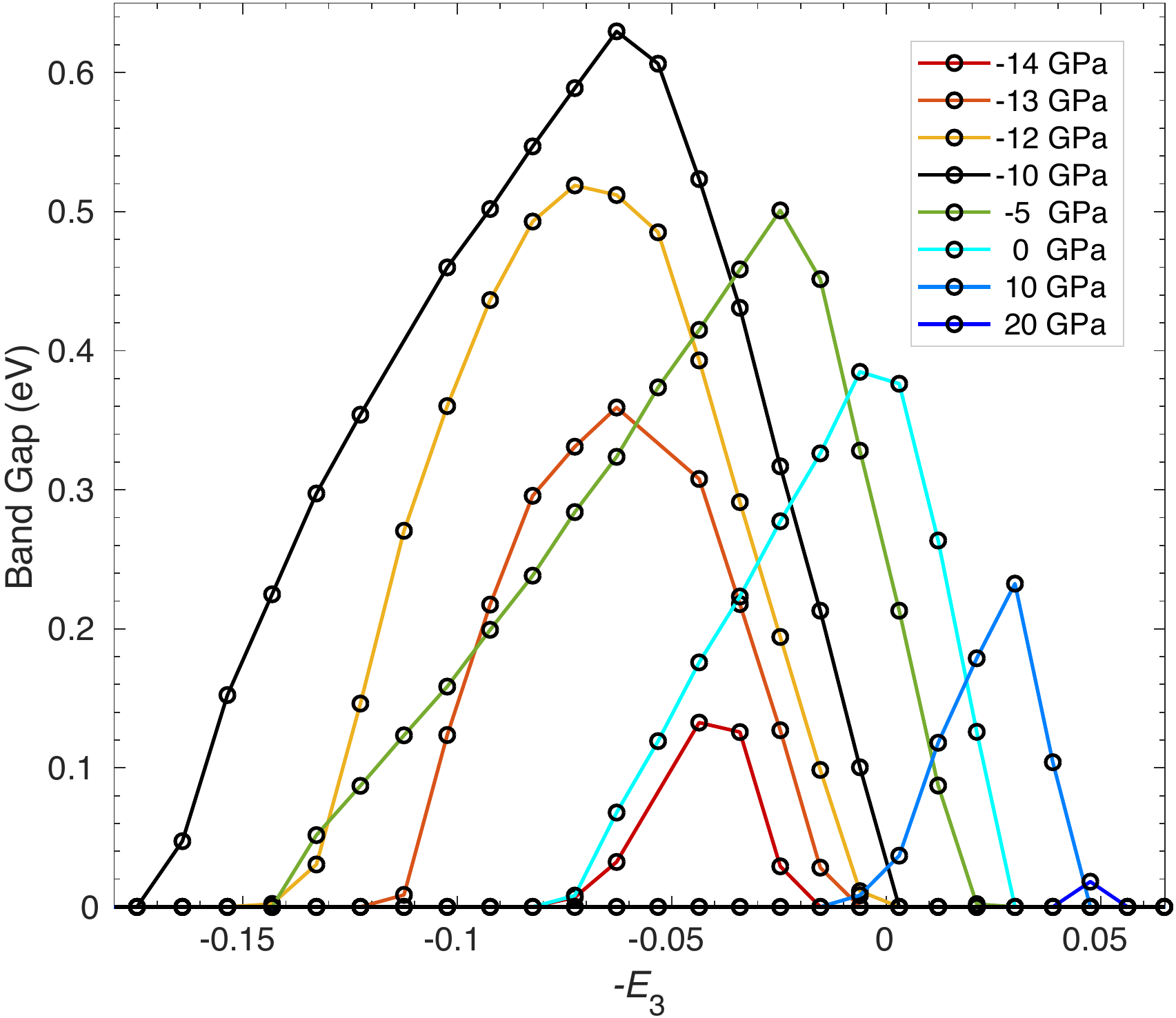}  
    \caption{Band gap width in deformed Si I vs. strain $E_3$ at various fixed $\sg_1=\sg_2$, ranging from $-14$ to $+20 \,$GPa.
}
    \label{Fig6gap}
\end{figure}

The electronic structure in Si I had been studied under different combinations of $\sg_3$ and fixed $\sg_1=\sg_2$.
Examples of the electronic band gap vs. compressive or tensile strain are given in Fig. \ref{Fig6gap}.
 For each $\sg_1=\sg_2$, there is a strain $E_3$, for which the band gap reaches its maximum,
 while a substantial deformation in any direction reduces the band gap to zero
 beyond the metallization curve, shown in Fig.~\ref{Fig5}.
 The band gap is  maximal near $\sg_1=\sg_2=\sg_3 \approx -10\,$GPa, see Fig.~\ref{Fig6gap}.

The Si I $\rightarrow$ Si II PT is accompanied by metallization -- an electronic transition from semiconducting to metallic phase.
However, the relation between the structural and electronic properties was not established.
We found that the electronic transition precedes the structural PT for all combinations of stresses:
a sufficiently deformed Si I under stress is metallic, see Fig.~\ref{Fig6gap}.
Also, this electronic transition does not change the continuity   of the stress-strain curves and their first derivatives (Fig. \ref{Fig5});
this differs from the stress discontinuity in magneto-structural phase transitions \cite{JChemPhys143n6p064707y2015}. 
The metallization curve is closed in the $(\sg_3,\sg_1)$ plane;  it can be approximated by two straight lines
$\sg_{3m}= -5.605 + 0.8417 \sg_1$
and $\sg_{3m}= 13.04 + 1.396 \sg_1$, and a parabolic section $\sg_{3m}= 11.95 + 2.378\sg_1+0.16 \sg_1^2$.
Metallization can be caused by compressive and tensile stresses (or their combination).
{While one of the metallization lines is relatively close and approximately parallel to the Si I$\rightarrow$Si II PT line,
the semiconducting (non-metallic) region is compact and its closed boundary surrounds the stress-free Si I.
According to Fig.~\ref{Fig5}, metallization occurs deeply in the region of stability of Si I. }
Under hydrostatic pressure, metallization occurs at compressive 36.82 GPa and tensile 13.91 GPa.
Under uniaxial loading, metallization is at compressive 5.4 GPa and tensile 12.78 GPa, i.e. the effect of non-hydrostatic stresses is extremely strong.
Under biaxial loading at $\sg_3=0$, the electronic transition happens at compressive 6.69 GPa and tensile 8.792 GPa.
Intersection of the metallization curve with the line for elastic instability of Si II in the stress plane in Fig. \ref{Fig5} does not have any meaning because strains for metallization correspond to the region of stability of Si I (compare Figs. \ref{Fig3} and \ref{Fig6gap}).


\section{\label{INSTABILITY3}ELASTIC LATTICE INSTABILITY UNDER TRIAXIAL LOADING}

\begin{figure}[ht]
    \centering
    \includegraphics[width=0.48\textwidth]{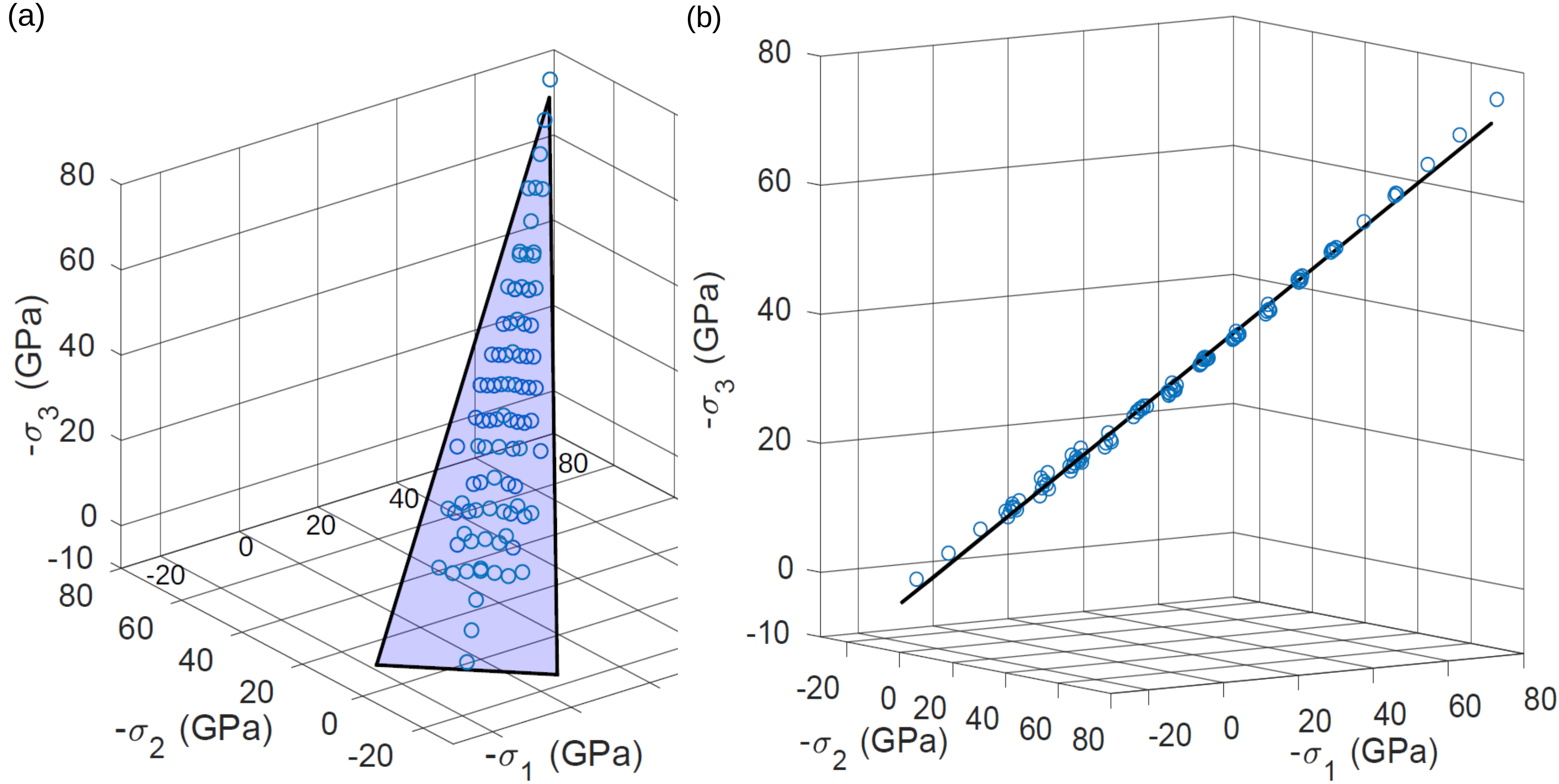}
    \caption{{Criterion for Si I$\rightarrow$Si II PT in the space of triaxial stresses.} (a) DFT results (points) lie in a plane with good accuracy (best fit),  giving a constant value of modified transformation work in Eq.~\ref{l-98-10-0ac1++}. (b) Result in (a) rotated to visualize an approximate plane.   DFT data is very close to the modified transformation work plane.
   }
    \label{Fig7}
\end{figure}

Evidently, DFT results for $\sg_1 \neq \sg_2$ case in Fig.~\ref{Fig7}
 suggest that the criterion for forward Si I$\rightarrow$Si II PT can be described accurately in 3D space of normal stresses by a plane
\bey
\sg_3=- 9.911+ 0.4145 (\sg_1+ \sg_2) . \quad
\label{eshelbi-98-08}
\eey
It is very surprising that the elastic instability for a material with strong physical and geometric nonlinearities can be approximated by a linear criterion.


\section{\label{PHASEFIELD}LATTICE INSTABILITY UNDER STRESS TENSOR: THE PHASE FIELD APPROACH}
As  shown in  \cite{Levitas-IJP-13,Levitasetal-Instab-17,Levitasetal-PRL-17,Levitas-IJP-18}, a PT condition linear in normal stress  can be derived within the phase field approach to martensitic PTs.  Using several steps and assumptions, the following instability criterion for  Si I$\rightarrow$Si II PT  was derived:
\bey
2 W=\fsg
 {\fg :} {\fg F}_{e}^{{\ssst T}-1}  \cdot
\frac{d^2
\bar{\fg U}_t
 }{d  \eta^2}\Big|_{\eta=0}  {\fg \cdot} {\fg F}^{\ssst T}_e
\geq
 2  A,
\label{l-98-10-0ac1++}
\eey
where deformation gradient  $\bar{\fg U}_t(\eta) \equiv \fg I+ \bar{\fvep}_t (\eta)$, and other material parameters
(e.g., thermal energy, elastic moduli, and transformation strain $\bar{\fvep}_t (\eta)$) depend on the order parameter $\eta$,
which changes during the transformation process from $\eta=0 $ for Si I
[i.e.,  $\bar{\fvep}_t(0)=\fg 0$] to $\eta=1$ for Si II [i.e.,  $\bar{\fvep}_t(1)={\fvep}_t = diag(   \vep_{t1},   \vep_{t1},    \vep_{t3}) $].
$W$ is called  the modified transformation work  \cite{Levitasetal-Instab-17}, and $A$ is the magnitude of the double-well barrier.
{  For cubic to tetragonal transformation,
$ \frac{d^2 \bar{\fg U}_t   }{d  \eta^2}\Big|_{\eta=0}=2  diag(b_1  \vep_{t1}, b_1  \vep_{t1}, b_3  \vep_{t3}) $,
where $b_i$ are the coefficients in the interpolation of $\bar{\fvep}_t (\eta)$.
For the loading by three stresses normal to the chosen above faces, all tensors in Eq.~\ref{l-98-10-0ac1++} are coaxial, tensors ${\fg F}_{e}^{{\ssst T}-1}  $ and ${\fg F}_{e}^{{\ssst T}}  $ eliminate each other,
and  Eq.~\ref{l-98-10-0ac1++} reduces to the linear modified transformation work criterion:
\bey
 W=     b_3 \sg_3 \vep_{t3} +  b_1  (\sg_1+ \sg_2)   \vep_{t1} =  A.
 \label{l-98-10-0ac}
\eey
{The equality is used to describe combination of stresses at the limit of stability and calibrate material parameters.
  $W$  reduces to the transformation work for $b_1=b_3=1$.} The consequence of Eq.~\ref{l-98-10-0ac} for cubic-to-tetragonal PT is that, with $\vep_{t1}=\vep_{t2}$, the stresses $\sg_1$ and $\sg_2$ contribute to the instability criterion via $\sg_1+ \sg_2$, i.e., as in Eq.~\ref{eshelbi-98-08}.
Comparing Eqs.~\ref{l-98-10-0ac} and \ref{eshelbi-98-08} with $\vep_{t1}=U_{t1}-1=0.243$  and $\vep_{t3}=U_{t3}-1=-0.514$ leads to  $ A({\theta})/b_3= 5.094\,$GPa   and $b_3/b_1=1.141$.

{\par}  When shear stresses $\tau_{ij}$ are applied, causing nonzero deformation gradients $F_{21}$, $F_{31}$, $F_{32}$,
with rigid-body rotations  excluded by imposing a constraint $F_{12}=F_{13}=F_{23}=0$, Eq.~\ref{l-98-10-0ac1++}
reduces to
\begin{eqnarray}
&& W= b_3 \sg_3 \vep_{t3} +  b_1  (\sg_1+ \sg_2)   \vep_{t1} \, + \hfill \hspace{9mm}
 \label{W2}\\
\hfill  &&  \,    \frac{b_1\vep_{t1}-b_3 \vep_{t3}}{F_{11}^e F_{22}^e}\left[ \tau_{32}F_{32}^e F_{11}^e + \tau_{31}( F_{31}^e F_{22}^e -  F_{32}^e F_{21}^e)  \right] =
  A,  \quad
\nonumber
\end{eqnarray}
where $(b_1\vep_{t1}-b_3 \vep_{t3})/A=0.143$ and the terms proportional to 
 $\vep_{t2}-\vep_{t1}$ are eliminated.
With  transformation shears absent in a cubic-to-tetragonal PT, the shear transformation work is absent.
The terms proportional to the shear stresses are due to geometric nonlinearity (finite strains); they do not contain any additional material parameters.
Shear stresses change geometry of the crystal, and this affects transformation work along the normal components of transformation strain.

{\par} Note that Eq.~\ref{W2} is not invariant under exchange $1 \! \leftrightarrow \! 2$ because of imposed kinematic constraint.   For the obtained parameters, and because $F^e_{ii}>0$ and $ \tau_{ij}F_{ij}^e>0$,
 when $\tau_{32} $ and $F_{32}^e$ or $\tau_{31} $ and $F_{31}^e$ are applied alone, contribution of shear stresses to $W$ is positive, i.e., they promote tetragonal instabilities.
 Shear stress $\tau_{21} $ (more exactly, elastic shear strain $F^e_{21}$) alone or with $\tau_{32} $ does not contribute to the instability  condition;
but  $\tau_{21} $ contributes when two other stresses, $\tau_{31} $ and $\tau_{32} $, are applied simultaneously, and depending on signs of all shear stresses, $\tau_{21} $ may promote or suppress tetragonal instability.

\section{\label{SHEAR1}SHEAR STRESS-STRAIN CURVES AND SHEAR  LATTICE INSTABILITY}
Increasing simple shears   $F_{21}$, $F_{31}$, $F_{32}$ and their combinations were applied at various fixed $F_{11}=F_{22}$ ($2-3 \%$ before and after tetragonal instability points) and $F_{33}$, for which stresses $\sg_1=\sg_2$ were equal to values given in Fig. \ref{Fig8}  before shear loading.
Typical shear stress $\tau_{31}$  -- deformation gradient $F_{31}$ curves are shown in Fig. \ref{Fig8}.
Shear instability starts at the maximum shear stress.
This instability does not lead to Si II but rather to possible amorphization or hexagonal diamond Si IV (which is beyond our present focus). 
Here we study the effect of shear stresses on the tetragonal instability mode, responsible for the PT to Si II, which typically happens  before the shear instability is reached.

\begin{figure*}[t]
    \centering
   \includegraphics[width=0.8\textwidth]{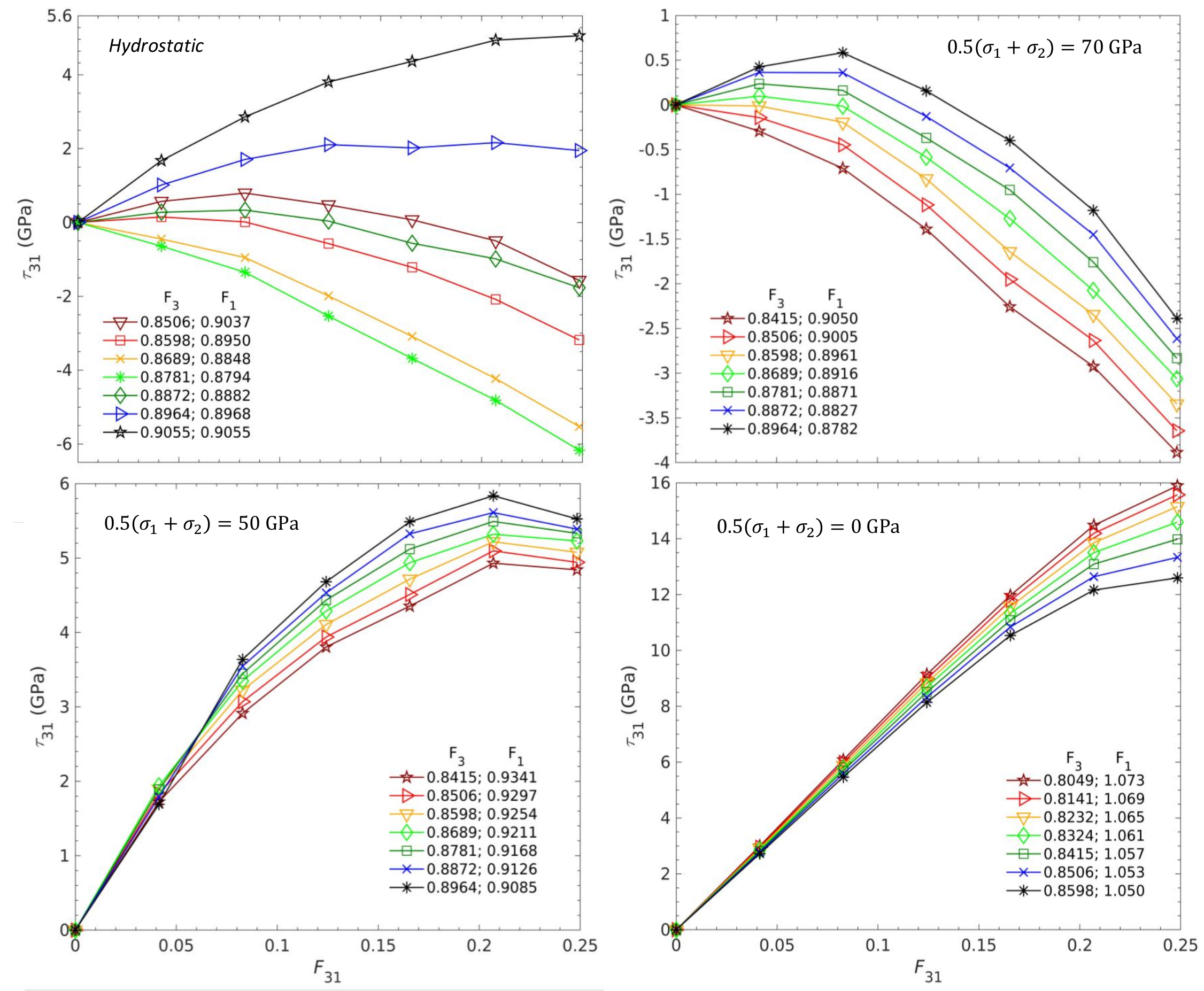}
    \caption{{  Shear stress $\tau_{31}$  - deformation gradient $F_{31}$ curves at
various fixed $F_{1}=F_{2}$ ($2-3 \%$ before and after tetragonal instability points) and $F_{3}$.}
Plots are for initially hydrostatic loading and for loading with  stresses $\sg_1=\sg_2$,
initially (before shear loading) equal to the shown values. Due to geometric nonlinearity, normal stresses vary with increasing shears but $0.5 (\sg_1+\sg_2)\simeq const$.
Legends tabulate values of $F_{3}$ and $F_{1}$: the middle (green) corresponds to the tetragonal instability without shear stresses,
while the three values below (above) correspond to the Si I before (after) tetragonal instability point.
   }
       \label{Fig8}
\end{figure*}

{\par} Under an initial (before shear) hydrostatic compression, shear stresses for any $F_{31}$ in the cubic phase reduce with increasing volumetric strain and pressure
{ (see curves for four lower combinations of $F_{3}$ and $F_{1}$ in Fig.~\ref{Fig8}~(a)), } which is qualitatively consistent with the limited results in \cite{Umeno-Cerny-PRB-08} for the $[11\bar{2}](111)$ slip system.
After reaching the instability pressure for Si I $\rightarrow$ Si II PT and following the tetragonal branch of deformation gradient
(see curves for three upper combinations of $F_{3}$ and $F_{1}$ in Fig.~\ref{Fig8}~(a)), a crossover is observed and a shear stress for any $F_{31}$ increases with further growth of $F_3$ and volumetric strain,  while the pressure reduces along the unstable branch of pressure -- $E_3$ (or volumetric strain) curve.
The shear instability at an infinitesimal shear starts at 72 GPa, i.e., below the tetragonal mode of lattice instability.
This may explain amorphization in nanocrystalline Si I under increasing pressure when PT to Si II is kinetically suppressed \cite{Debetal-Nat-01}.
Amorphization may be caused by virtual melting \cite{Levitas-PRL-2005} after crossing metastable continuation of the melting line, since melting temperature for Si reduces with pressure.

{ The effect of pressure on the $\tau_{21}-F_{21} $ curves is qualitatively similar to that for the $\tau_{31}-F_{31} $ curves. However, a shear instability for any $F_{21} \leq 0.25$ starts after the tetragonal instability. }

{\par}
At a non-hydrostatic initial loading,
physics is essentially different.
At the initial stress $\sg_1=\sg_2=- 69.61 \,$GPa,
shear instability for an infinitesimal $F_{31}$ starts practically simultaneously with the  tetragonal instability
(see green curve for the middle values of $F_{3}$ and $F_{1}$ in Fig.~\ref{Fig8}~(b)).
Before tetragonal instability, the shear instability shifts to larger shears.  Shear stress $\tau_{31}$ decreases with increasing $|E_{3}|$ monotonously, in contrast to hydrostatic loading. { At the same time, shear instability
occurs for $F_{21} > 0.2$ after tetragonal instability.} Both $\tau_{31}$ and $\tau_{21}$ decrease with increasing $|E_{3}|$; { for all  $|E_{3}|$  and equal shears one has $\tau_{31}<\tau_{21}$.}
This tendency in stress-strain curves is kept to  $\sg_1=\sg_2=- 39.63 \,$GPa  with increasing   shear instability strain  $F_{31}$ and without essential change in the instability strain  $F_{21}$. Amplitude  of both shear stresses increases  with reducing $|\sg_1|=|\sg_2|$. Effect of lateral $F_1$ and
corresponding axial $F_3$ compressions on both shear stress-strain curves reduces with decreasing $|\sg_1|=|\sg_2|$. 
At   $|\sg_1|=|\sg_2|=29.68 \,$GPa a crossover occurs, i.e., shear stresses slightly increase with  $|E_{3}|$.

\begin{figure*}[t]
  \includegraphics[width=0.8\textwidth]{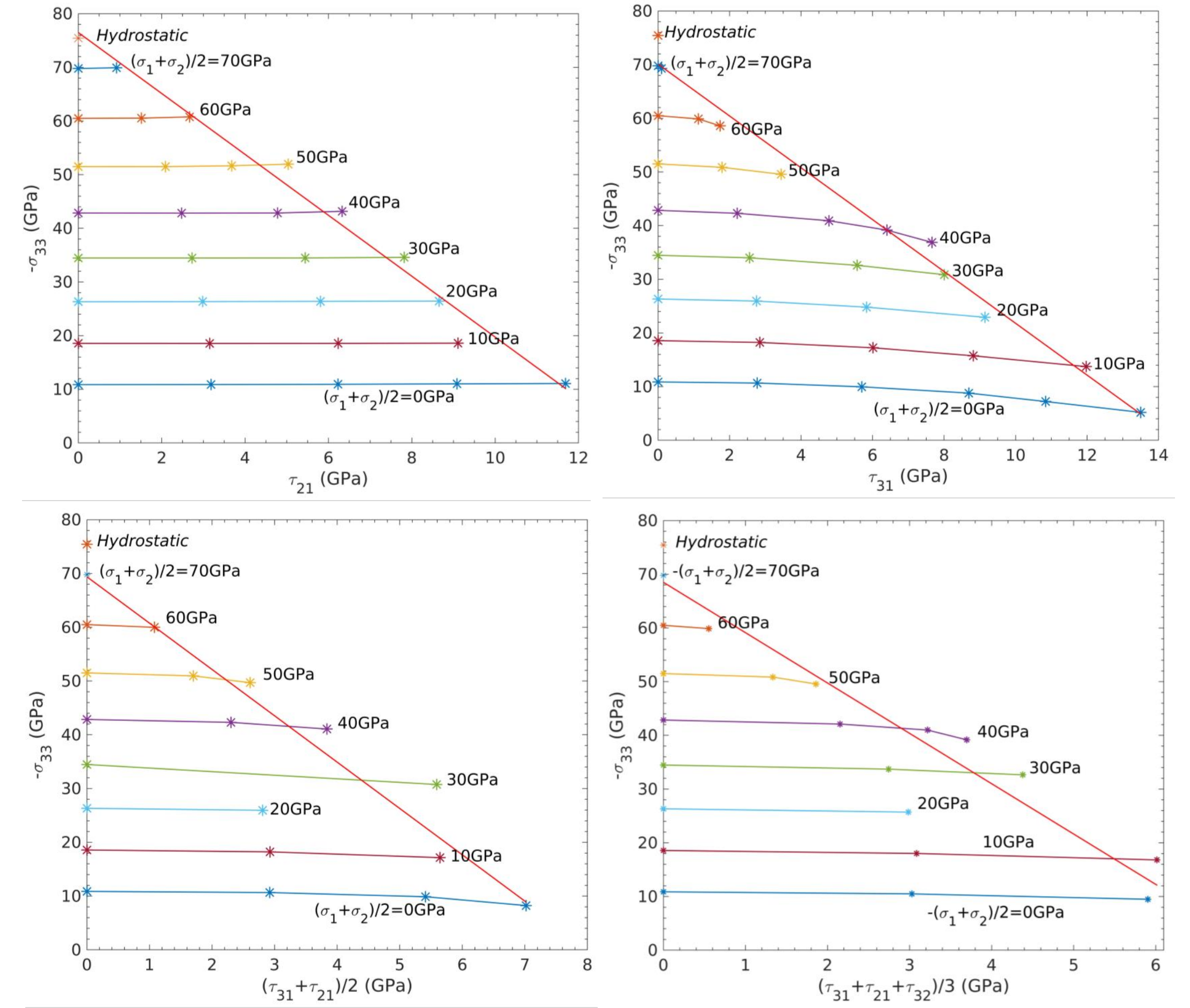}
    \caption{
    The effect of various combinations of shear stresses on the tetragonal instability stress $\sg_3$ for different $\sg_1=\sg_2$.
Points with the largest shear stress approximately correspond to shear instability.
Straight inclined lines are linear approximations of the relationship between $\sg_3$ and shear stresses for shear instability.
       }
       \label{Fig9}
\end{figure*}

\section{\label{SHEAR}EFFECT OF SHEAR STRESSES ON TETRAGONAL INSTABILITY}

 {
The $\sg_3-E_3$ curves [along the path in the $(F_1 = F_2, F_3)$ plane corresponding to $\sg_1=\sg_2$ before shear] have been obtained for different fixed shears.
The instability stress in Fig. \ref{Fig9} is determined as the local maximum of $|\sg_3|$, see Fig.~\ref{Fig4}.
While during shear $\sg_1 \neq \sg_2$ but their sum $\sg_1+\sg_2$ practically does not change.
 That is why  curves  in Fig. \ref{Fig9} are given for the approximately fixed values of  $(\sg_1+\sg_2)/2$.

 {\par}  In addition, absolute and relative deviations between the actual instability stress $\sg_3$ and   $\sg_3^{an}$ based on the analytical prediction  (\ref{W2}) are presented in Figs.~\ref{Fig10} and \ref{Fig11}.
As we already discussed (see linear Eq.~\ref{l-98-10-0ac}), shear stress $\tau_{21}$ alone does not contribute to the analytical instability condition (\ref{W2})
and practically (within the  relative error of $6 \%$) does not affect the  instability stress $\sg_3$  in a  broad range of shear stresses $\tau_{21}$ below the shear instability,
which is approximately described by $\tau_{21}^{in}= 11.09 + 0.1470 \, \sg_1$.
    Shear instability stress $\tau_{21}^{in}$ varies from 11.09 GPa at $\sg_1=0$ to 0 at $\sg_1\simeq -75.44 \,$GPa.

{\par}
 An increasing shear stress $\tau_{31}$ causes  some reduction in the instability stress $\sg_3$ 
 (Fig.~\ref{Fig9}).
The relative error of the instability stress with respect to the analytical prediction (\ref{W2})  for most combinations of
of $\tau_{31}$ and $(\sg_1+\sg_2)/2$ is between $+4\%$ and $-6\%$.
However, there are three outliers at a large shear stress $\tau_{31}>8.5 \,$GPa.
At these points, stresses $-(\sg_1+\sg_2)/2$ are small (from  $-2$ to $10\,$GPa) and the corresponding instability stress $-\sg_3$  is also small (10--$18 \,$GPa);
a ratio of smaller numbers with finite absolute errors has a larger relative error.
The absolute error $\sg_3-\sg_3^{an}$ for these points is just within $\pm 1 \,$GPa, see Fig.~\ref{Fig11}.
A larger error of $\pm 2 \,$GPa  appears for  small shear stresses but  large $-\sg_1$ and consequently $-\sg_3$ (from $50 $ to $75 \,$GPa), i.e., close to the shear instability. A relative error there remains within  $\pm 4 \%$.

Thus, the main effect of a shear stress  $\tau_{31}$ on the instability stress $\sg_3$ is due to the theoretically predicted  geometric nonlinearity with zero linear term.
The combined effect of two and three shear stresses on the instability stress $\sg_3$ is smaller than the effect of $\tau_{31}$ alone
(a) because of a smaller averaged shear stress that causes shear instability and
(b) because of a small contribution of $\tau_{21}$ for two shear stresses and opposite contribution of $\tau_{21}$ for three stresses, according to Eq.~(\ref{W2})
 for all positive shear stresses. Deviation from the prediction (\ref{W2})  does not exceed $\pm 4 \%$.

 Thus, the tetragonal lattice instability under action of all six components of the stress tensor can be described by the  critical value of the modified transformation work, namely,
 by Eq.~(\ref{W2}), which
(a) is linear in normal stresses, depends on $\sg_1+\sg_2$, and has only two adjustable  coefficients ($b_1$ and $b_3$);
(b) is independent of $\sg_1 - \sg_2$  and shear stress $\tau_{21}$ acting alone or with one more shear stress;
(c) contains a geometric nonlinear term describing contribution of all shear stresses without any additional adjustable parameters.

{\par} For a neglected effect of shear stresses, an absolute deviation of $\sg_3$ from the linear expression (\ref{l-98-10-0ac})
is within $2 \,$GPa for  $\tau_{31}< 5 \,$GPa and within $3 \,$GPa   for  $\tau_{31}< 8 \,$GPa,
while its relative deviation is within $10 \%$ for    $\tau_{31}< 8 \,$GPa.
}

 \begin{figure}[t]
    \includegraphics[width=0.4\textwidth]{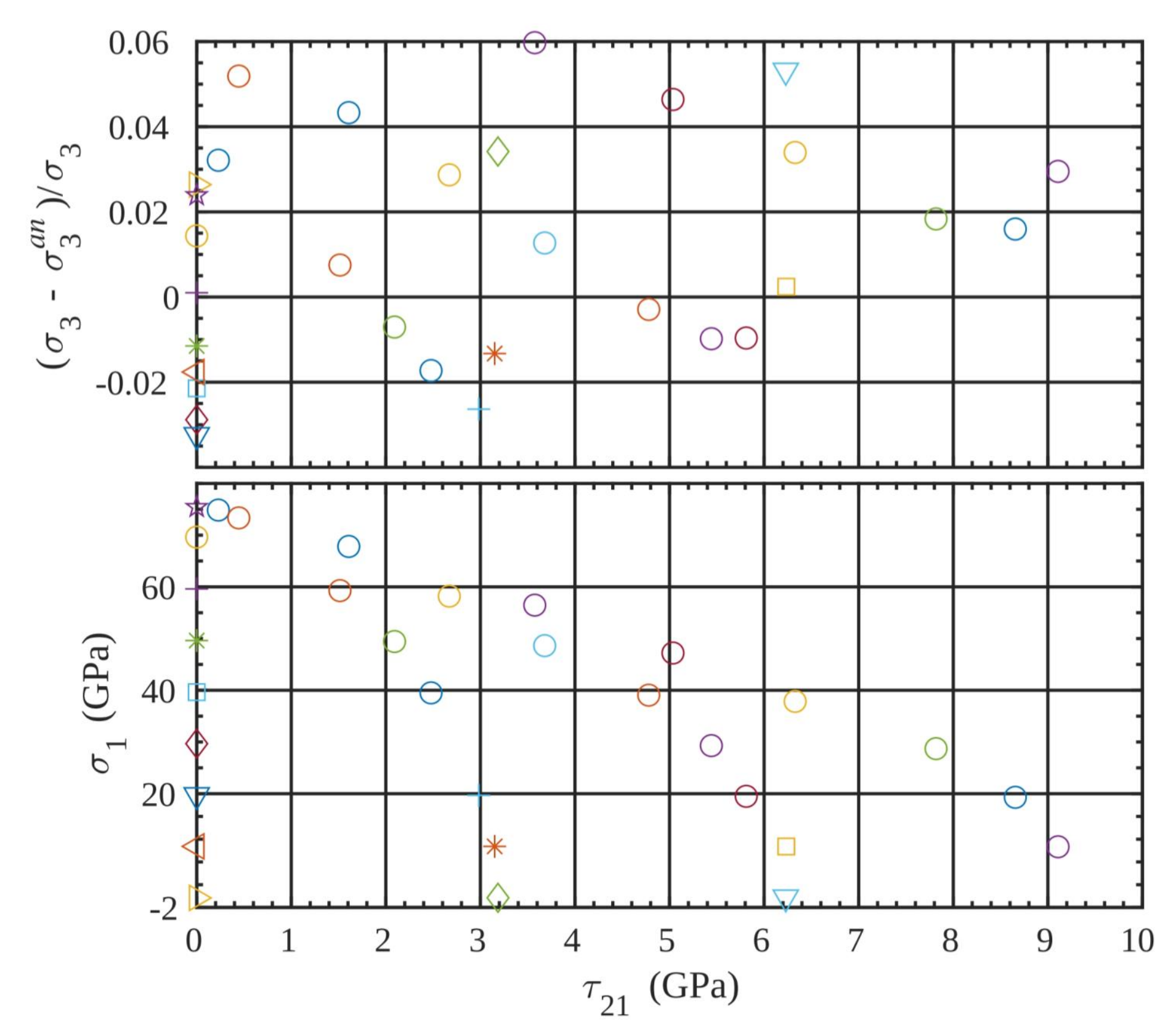}
    \caption{{ Relative difference between the actual instability stress $\sg_3$ and the instability stress $\sg_3^{an}$ based
    on analytical prediction  Eq.(\ref{W2}) and corresponding values of $-(\sg_1+\sg_2)/2$ versus shear stress    $\tau_{21}$.
       }}
       \label{Fig10}
\end{figure}

 \begin{figure}[t]
    \includegraphics[width=0.35\textwidth]{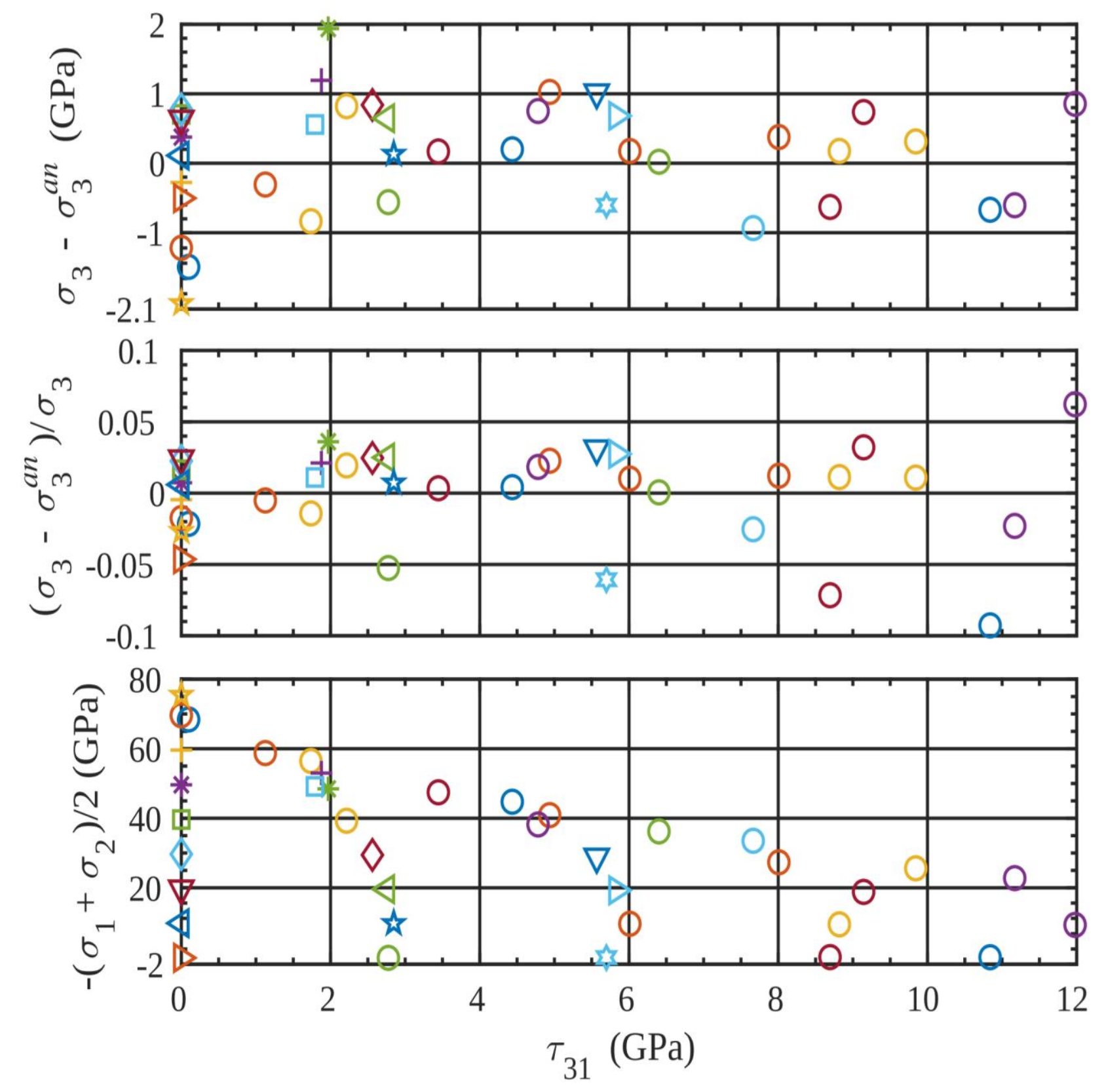}
    \caption{{ Absolute and relative difference between the actual instability stress $\sg_3$ and the instability stress $\sg_3^{an}$ based
    on analytical prediction  Eq.(\ref{W2}) and corresponding values of $-(\sg_1+\sg_2)/2$  versus shear stress  $\tau_{31}$.
       }}
       \label{Fig11}
\end{figure}

\section{\label{CONCLUDING} SUMMARY}
We augment standard $T$-$P$ equilibrium diagram with the criteria for  structural and electronic instabilities under controlled and predictable non-hydrostatic deformation, providing guidance for creating more accessible synthetic processing routes for new or known high-hydrostatic-pressure phases, including those with novel properties.
To exemplify this pivotal concept, we performed a comprehensive DFT study of the PT between semiconducting Si I and metallic Si II under the application of a general stress tensor, and  investigated the stress-strain curves, elastic lattice instabilities, and metallization.
The PT pressure under hydrostatic condition is $\approx \! 20$ times larger than under uniaxial loading.
Such a strong effect of nonhydrostaticity  at least partially explains  the significant difference between the experimental PT pressure (9-12 GPa) and  the instability pressure of 75.81~GPa, as well as a scatter in the experimental data under quasi-hydrostatic conditions.

{\par} Although the stress-strain curves and their first derivatives are continuous, metallization precedes the structural PT.
That means that under stresses there is a metallic Si I.
Metallization can be caused by compressive or tensile stresses, and the effect of non-hydrostatic stresses is very strong.
In the $(\sg_1,\sg_3)$ plane in Fig.~\ref{Fig5}, it is described by a closed contour (given  roughly by two straight lines and a parabolic cap).
Only one of the metallization lines is relatively close and approximately parallel to the Si I$\rightarrow$Si II PT line.
Notably,  along other two lines, metallization occurs deeply in the region of stability of Si I and is not causing the Si I$\rightarrow$Si II structural PT.

{\par} Our key result is that Si I $\rightarrow$ Si II PT can be  described by a critical value of the modified transformation work (Eq.~\ref{W2}), obtained from a phase field formalism.
From this, a PT criterion was found to be linear in normal stresses and the effect of shear stresses is described via a geometric nonlinear term (with no additional adjustable constants). The PT criterion contains just two adjustable parameters, determined  by the instability at two different normal stress states (no shears needed), straightforwardly obtained from DFT.  Thus, our criterion accurately describes the lattice instability for a general stress tensor, which can then be assessed for optimal and easiest processing route.

{\par} While Si I$\rightarrow$Si II PT occurs due to elastic instability, the modified transformation work criterion (\ref{l-98-10-0ac1++}), simplified in Eq.~(\ref{W2}), is based on completely different principles and assumptions.
In particular,  it considers the entire dissipative PT process described by the transformation strain tensor and does not include the terms with a discontinuity in elastic moduli, to avoid nonlinearity in normal stresses.
Using these paradoxical results, we formulated a problem of finding a fundamental relationship between the lattice instability and  modified transformation work criterion, which will be studied in the future.  The elastic instability analysis for the simplest model quadratic in $\fg E$ energy  \cite{Levitas-Si-MRL-17}  qualitatively reproduces our main results for relatively low stresses.

{\par} The present results are significant for creating new, more practically achievable and economic synthetic processing routes for  discovery and stabilizing materials with novel properties, as well as for advancing and calibrating large-strain phase field models, e.g., in  \cite{Levitas-IJP-13,Levitasetal-Instab-17}.
Competition of the instability stresses [rather than the relative enthalpy, or Gibbs free energy, minima] can serve as a basis  for phase selection.
Critically, the results enable ways to reduce PT pressure due to non-hydrostatic stresses and plastic strains by an order of magnitude or more \cite{levitas-prb-04,Blank-Estrin-2014,Ji-etal-12,Edalati-Horita-16}.
They can also be used for quantitative studies of the influence of crystal defects on phase transitions  \cite{Levitas-Mahdi-Nanoscale-14,Javanbakht-Levitas-PRB-16}, and   quantitatively rationalize connections between PT conditions for  ideal and real (defective) crystals.
\\
\\
{\bf  ACKNOWLEDGEMENTS}
\\
\\
For theory-guided synthesis, NAZ and DDJ are supported by the U.S. Department of Energy (DOE), Office of Science, Basic Energy Sciences, Materials Science and Engineering Division. NAZ completed this work under support from Advanced Manufacturing Office of the Office of Energy Efficiency and Renewable Energy through CaloriCool{\texttrademark}, focused on applications to first-order phase transitions in caloric materials.
Ames Laboratory is operated for DOE by Iowa State University under contract DE-AC02-07CH11358.
VIL and HC acknowledge support  from  NSF (CMMI-1536925 and DMR-1434613)  ARO (W911NF-17-1-0225), and ONR (N00014-16-1-2079). The MD simulations were performed at  Extreme Science and Engineering Discovery Environment (XSEDE),  allocations TG-MSS140033  and MSS170015.

\bibliography{Si}

\end{document}